\documentclass[prd,aps,nofootinbib,floats,letterpaper,floatfix,
superscriptaddress,groupedaddress,eqsecnum]{revtex4}
\usepackage[usenames]{color}
\usepackage{epsfig}
\usepackage{amssymb,amsmath}

\newcommand{\f}[2]{\frac{#1}{#2}}

\newcommand{\dps}{\displaystyle}

\begin{document}
\def\be{\begin{equation}}
\def\ee{\end{equation}}
\def\beq{\begin{eqnarray}}
\def\eeq{\end{eqnarray}}
\def\LL{\left[\left[}
\def\RR{\right]\right]}

\def\msun{M_\odot}

\def\nn{\nonumber}
\def\lappreq{\! \stackrel{\scriptscriptstyle <}{\scriptscriptstyle
\sim}\!}
\def\gappreq{\!\stackrel{\scriptscriptstyle >}{\scriptscriptstyle \sim}\!}
\def\ver{\vskip 12pt}

\def\ii{{\rm i}}   

\def\Y{Y_{l m}}
\def\YS{Y^*_{l m}}
\def\Yt{\f{\partial Y_{l m}}{\partial\theta }}
\def\Ytt{\f{\partial^2 Y_{l m}}{\partial\theta^2 }}
\def\Yf{\f{\partial Y_{l m}}{\partial\phi }}

\def\EM{e^{-i\omega t}}
\def\EP{e^{+i\omega t}}
\def\N{N_{l m}}
\def\T{T_{l m}}
\def\V{V_{l m}}
\def\L{L_{l m}}
\def\X{X_{l m}}
\def\G{G_{l m}}
\def\hh{h^{0}_{l m}}
\def\hs{h^{1}_{l m}}
\def\pps{\Psi_{l m}}                                                                      
\def\IL{\relax{\rm I\kern-.18em L}}         

\def\lll{\left[}
\def\rr{\right]}
\def\bl{\Biggl[}
\def\br{\Biggr]}
\def\gbl{\Biggl\{}
\def\gbr{\Biggr\}}
\def\o{\omega}

\title{Gravitational-wave signatures of the absence of an event
  horizon.\\II. Extreme mass ratio inspirals in the spacetime of a thin-shell
  gravastar}

\author{Paolo Pani} \email{paolo.pani@ca.infn.it} \affiliation{Dipartimento di
  Fisica, Universit\`a di Cagliari, and INFN sezione di Cagliari, Cittadella
  Universitaria 09042 Monserrato, Italy}

\author{Emanuele Berti} \email{berti@phy.olemiss.edu} \affiliation{Department
  of Physics and Astronomy, The University of Mississippi, University, MS
  38677-1848, USA} \affiliation{Theoretical Astrophysics 350-17, California
  Institute of Technology, Pasadena, CA 91125, USA}

\author{Vitor Cardoso} \email{vitor.cardoso@ist.utl.pt}
\affiliation{Department of Physics and Astronomy, The University of
  Mississippi, University, MS 38677-1848, USA} \affiliation{Centro
  Multidisciplinar de Astrof\'{\i}sica - CENTRA, Dept. de F\'{\i}sica,
  Instituto Superior T\'ecnico, Av. Rovisco Pais 1, 1049-001 Lisboa, Portugal}

\author{Yanbei Chen} \email{yanbei@tapir.caltech.edu} 
\affiliation{Theoretical Astrophysics 350-17, California Institute of
  Technology, Pasadena, CA 91125, USA}

\author{Richard Norte} \email{norte@caltech.edu} 
\affiliation{Theoretical Astrophysics 350-17, California Institute of
  Technology, Pasadena, CA 91125, USA}

\begin{abstract}
  We study gravitational-wave emission from the quasi-circular, extreme mass
  ratio inspiral of compact objects of mass $m_0$ into massive objects of mass
  $M\gg m_0$ whose external metric is identical to the Schwarzschild metric,
  except for the absence of an event horizon.  To be specific we consider one
  of the simplest realizations of such an object: a nonrotating thin-shell
  gravastar. The power radiated in gravitational waves during the inspiral
  shows distinctive peaks corresponding to the excitation of the polar
  oscillation modes of the gravastar.  For ultra-compact gravastars the
  frequency of these peaks depends mildly on the gravastar compactness. For
  masses $M\sim 10^6M_\odot$ the peaks typically lie within the optimal
  sensitivity bandwidth of LISA, potentially providing a unique signature of
  the horizonless nature of the central object. For relatively modest values
  of the gravastar compactness the radiated power has even more peculiar
  features, carrying the signature of the microscopic properties of the
  physical surface replacing the event horizon.
\end{abstract}

\maketitle

\section{Introduction}
\label{sec:intro}

One of the most elusive properties characterizing the black hole (BH)
solutions of general relativity is the presence of their event
horizon. Traditional electromagnetic astronomy can at best yield lower limits
on the gravitational redshift corresponding to hypothetical surfaces replacing
the event horizon (see
\cite{Narayan:2005ie,Psaltis:2008bb,Visser:2009xp,Abramowicz:2002vt} for
different viewpoints on this delicate issue). However, present and planned
gravitational-wave (GW) detectors offer new prospects for ``directly''
observing BHs and probing their structure \cite{Sathyaprakash:2009xs}.

From a gravitational point of view the structure of compact objects is encoded
in their free oscillation spectrum, i.e. in their quasinormal modes (QNMs)
\cite{Kokkotas:1999bd,Berti:2009kk}. In an effort to point out the peculiar
features of compact objects whose external metric is identical to the
Schwarzschild metric, but which do not possess an event horizon, in
Ref.~\cite{Pani:2009ss} (henceforth Paper I) we studied the free oscillations
of one of the simplest ultra-compact horizonless objects: a nonrotating
thin-shell gravastar. Our analysis completed previous investigations
\cite{Fiziev:2005ki,Chirenti:2007mk} by considering the thin shell as a
dynamical entity. The QNM spectrum of a thin-shell gravastar is complex and
profoundly different from that of a BH, mainly because of the different
boundary conditions at the surface replacing the event horizon.

As first proposed by Ryan \cite{Ryan:1995wh,Ryan:1997hg}, an exquisite map of
BH spacetimes can be constructed by observing the gravitational waveform
emitted when a small compact object spirals into the putative massive BH at
the center of a galaxy with the Laser Interferometer Space Antenna (LISA).  Li
and Lovelace refined the analysis showing that the small object's tidal
coupling encodes additional information about the metric of the spacetime
\cite{Li:2007qu}. In this paper we use the formalism developed in Paper I to
show that any surface replacing the BH event horizon will produce a very
specific signature in the gravitational signal emitted by the orbiting object
because of the resonant scattering of gravitational radiation, which can be
traced back to the different QNM spectrum of the two objects. In fact, here we
show that the QNMs of ultra-compact thin-shell gravastars can be excited
\emph{during the inspiral}, whereas Schwarzschild QNMs can only be excited by
particles plunging into the BH (see
Refs.~\cite{Kojima:1987tk,Gualtieri:2001cm,Pons:2001xs,Berti:2002ry} for a
discussion of the analogous problem of particles orbiting neutron stars).

This work is very similar in spirit to a previous study by Kesden and
collaborators \cite{Kesden:2004qx}. There are two main differences between our
work and theirs. The first difference is that Kesden {\it et al.} considered
boson stars rather than gravastars as BH strawmen, so no ``hard surface''
replaces the event horizon in their case. The second difference is that we
compute the radiation in a consistent perturbative framework, instead of using
``kludge'' waveforms that become increasingly inaccurate in the relativistic
regime. In this sense, this paper is the first ``strong-field'' calculation of
the potential gravitational signatures of inspirals into horizonless objects.

The plan of the paper is as follows. In Sec.~\ref{sec:grav_internal} we review
the equations describing axial and polar perturbations in the interior of a
gravastar and the matching conditions with the ordinary perturbations of a
Schwarzschild metric in the exterior spacetime. In Sec.~\ref{sec:grav_source}
we summarize the perturbed Einstein equations in the Bardeen-Press-Teukolsky
(BPT) formalism and we write down the source term appearing on the right-hand
side of the BPT equation for orbiting pointlike particles of mass $m_0\ll
M$. Then we discuss how the perturbation functions outside the shell (as
obtained by the matching conditions derived in Paper I) can be used to solve
numerically the BPT equation with a source given by the orbiting particle and
to compute the radiated power.  In Sec.~\ref{sec:results} we compare numerical
calculations of the power radiated by BHs and different gravastar models and
we stress potentially observable GW signatures of horizonless ultra-compact
objects. We conclude by discussing possible extensions of our work.

\section{Gravitational perturbations by a point particle}
\label{sec:perturbations}

The emission of gravitational waves by an extreme mass ratio binary system can
be computed by a perturbative approach. One of the two compact objects is
assumed to be an extended body of mass $M$, whose equilibrium structure is
described by an exact solution of the Einstein equations. The second object is
regarded as a point particle of mass $m_0\ll M$ perturbing the gravitational
field of the companion. This method has been applied to study gravitational
radiation from particles orbiting BHs (see
e.g.~\cite{Cutler:1993vq,Cutler:1994pb,Poisson:1995vs}) and neutron stars
\cite{Kojima:1987tk,Gualtieri:2001cm,Pons:2001xs}.  Here we apply the same
perturbative approach to compute the power radiated by particles orbiting a
thin-shell gravastar.

\subsection{Gravitational perturbations of the internal structure}
\label{sec:grav_internal}

In this and the following sections we shall use the same notation as in Paper
I. We consider a static thin-shell gravastar with metric
\cite{Visser:2003ge,Chirenti:2007mk}
\begin{equation} 
ds_0^2 = - f(r) dt^2+\frac{1}{h(r)}dr^2 + r^2(d\theta^2+\sin^2\theta d\varphi^2)\,,\label{eq:g0}
\end{equation}
with
\begin{equation}
f(r) = \left\{
\begin{array}{ll}
\displaystyle
h= 
1-\frac{2M}{r} \,, & r>a \,,\\
\\
\displaystyle 
\alpha\,h=
\alpha
\left(
1-\frac{8\pi\rho}{3}r^2
\right)
\,, & r<a \,,
\end{array}
\right.
\label{fr}
\end{equation}
where $M$ is the gravastar mass and $\rho=3M/(4\pi a^3)$ is the ``energy
density'' of the interior region. The junction conditions require the induced
metric to be continuous across the shell at $r=a$. This implies that $f(r)$
must also be continuous at $r=a$, i.e. $\alpha=1$. In order to compute the
radiation emitted by a gravastar perturbed by a massive point particle we must
compute the gravitational perturbations both inside and outside the
gravastar. Perturbations in the interior have been discussed in Paper I using
the Regge-Wheeler (RW) gauge. Here we only recall some results that are useful
to compute the power radiated by orbiting particles.

In the de Sitter interior, {\it both} axial and polar perturbations can be
reduced to the study of the master equation
\be
\frac{d^2 \Psi^{\rm in}}{d r_*^2}+\left [\omega^2-\frac{l(l+1)}{r^2}\,f(r)\right ]\Psi^{\rm in}=0\,,
\label{eq:masterrstar}
\ee
where the tortoise coordinate $r_*$ is defined by $dr/dr_*=f(r)$ and $f(r)$ is
given by Eq.~(\ref{fr}) with $r<a$. The regular solution at the center ($r=0$)
is
\be
\Psi^{\rm in}=
r^{l+1}(1-C(r/2M)^2)^{-i\frac{2M\omega}{2\sqrt{C}}}F
\left(\frac{l+2-i\frac{2M\omega}{\sqrt{C}}}{2},
\frac{1+l-i\frac{2M\omega}{\sqrt{C}}}{2},l+\frac{3}{2},C(r/2M)^2\right)\,,
\label{eq:interior_sol}
\ee
where $C\equiv(2M/a)^3=8\mu^3$ and $F(a,b,c,z)$ is the hypergeometric function
\cite{Abramowitz:1970as}. From $\Psi^{\rm in}$ and its derivative we can
obtain the Zerilli and RW perturbation functions as explained in Paper I.

For a thin-shell gravastar the background surface energy vanishes, $\Sigma=0$,  but the
surface stress-energy tensor $\Theta$ is, in general, nonvanishing.
This implies that the perturbation functions are discontinuous across the
shell. In Paper I we derived the matching conditions relating interior and
exterior perturbations in the RW gauge. For axial perturbations these matching
conditions read
\be 
\LL h_0\RR=0\,,\quad \LL\sqrt{h} h_1\RR=0\,,
\label{junctionaxial}
\ee
where $\LL\dots\RR$ denotes the ``jump'' of a given quantity across the shell,
i.e. the difference between the limits of the corresponding quantity as $r\to
a_\pm$. For a thin-shell gravastar Eqs.~(\ref{junctionaxial}) imply continuity
of the RW function and its derivative across the shell.
The treatment of polar perturbations is more involved and it yields the
following relations for the jump of the polar metric functions across the
shell:
\beq 
&&[[K]]=0\,,\quad [[K']]=-8\pi\frac{\delta\Sigma}{\sqrt{f(a)}}\,,\nonumber\\
&&\frac{2M}{a^2}\LL
H\RR-[[H\,f']]-2f(a)[[H']]+4i\omega[[H_1]]=16\pi\sqrt{f(a)}(1+2v_s^2)\delta\Sigma
\label{junctionpolar}\,.
\eeq
The parameter $v_s$ depends on the equation of state (EOS) on the thin shell,
$\Theta=\Theta(\Sigma)$:
\be 
v_s^2\equiv-\left(
\frac{\partial\Theta}{\partial\Sigma}
\right)_{\Sigma=0}\,,\label{eq:v}
\ee
and it has the dimensions of a velocity. A microscopic model of matter on the
thin shell is needed for a microphysical interpretation of $v_s$, but (roughly
speaking) this parameter is related to the sound speed on the shell.  We shall
follow the treatment in Paper I and treat $v_s$ as a free parameter, although
we will primarily focus on the (presumably more physical) range $0<v_s^2<1$.

\subsection{The source term and the BPT formalism}
\label{sec:grav_source}

A detailed treatment of the perturbative approach used to compute the
gravitational emission by a particle orbiting a polytropic neutron star can be
found in
Refs.~\cite{Gualtieri:2001cm,Pons:2001xs,Berti:2002ry,Berti:2002zz}. Here we
review the method with an emphasis on the modifications required to deal with
thin-shell gravastars.

The radial part $\pps (\omega, r)$ of the perturbation of the Weyl scalar
$\delta\Psi_4$ is defined as
\be
\label{psiquattro}
\pps (\omega, r)=\frac{1}{2\pi}
\int
d\Omega~dt~_{-2}S^\ast_{l m}(\theta,\phi)
\left[ r^4~\delta\Psi_4(t,r,\theta,\phi)\right]e^{ i\omega t}\,,
\ee
where $_{-2}S_{l m}(\theta,\phi)$ is a spin-weighted spherical harmonic of
spin $-2$.  The function (\ref{psiquattro}) can be expressed in terms of the Zerilli and RW
perturbation functions ($Z_{l}(\omega,r)$ and $Y_{l}(\omega,r)$, respectively)
as follows:
\beq
\label{relazione1}
\pps(\omega,r)&=&\frac{r^3\sqrt{n\left(n+1\right)}}{4\omega}\left[
V^{\rm ax} Y_l  +\left(W^{\rm ax} +2i\omega\right)\Lambda_+ Y_{l}  \right]\\
\nn
&-&\frac{r^3\sqrt{n\left(n+1\right)}}{4}\left[
V^{\rm pol} Z_{l}+\left(W^{\rm pol}+2i\omega\right)\Lambda_+  Z_{l}
\right]\,, \eeq
where $2n=(l-1)(l+2)$, $\Lambda_+={d/dr_*}+i\omega=
r^{-2}\Delta{d/dr}+i\omega$ and
\beq
W^{\rm ax}&=&\frac{2}{r^2}(r-3M)\,,\\
W^{\rm pol}&=&2\frac{nr^2-3Mnr-3M^2}{r^2(nr+3M)}\,.
\eeq
The functions $V^{\rm pol}$ and $V^{\rm ax}$ are the well-known Zerilli and RW
potentials:
\beq
V^{\rm ax}(r)&&=f\,\left(\frac{l(l+1)}{r^2}-\frac{6M}{r^3}\right )\,,\\
V^{\rm pol}(r)&&=\frac{2f}{r^3}\left(\frac{9M^3+3\lambda^2Mr^2+\lambda^2(1+\lambda)r^3+9M^2\lambda r}{
(3M+\lambda r)^2 } \right)\,,
\eeq 
with $\lambda=l(l+1)/2-1$.

The radial part of $\delta\Psi_4$ outside the shell can be computed from the
matching conditions discussed in Section~\ref{sec:grav_internal} and used as a
boundary condition for the integration of the inhomogeneous BPT equation
\cite{Bardeen:1973xb,Teukolsky:1973ha}
\be
\label{teukolsky}
{\cal L}_{BPT}\pps (\o, r)\equiv
\left\{\Delta^2\frac{d}{dr}\left[\frac{1}{\Delta}\frac{d}{dr}\right]+
\left[\frac{\left(r^4\omega^2+4i(r-M)r^2\o\right)}{\Delta}
-8i\o r-2n\right]\right\}\pps (\o, r)= -T_{l m}(\o, r),
\ee
where $\Delta=r^2-2Mr$ and the source term $T_{l m}(\o, r)$ describes the
point mass $m_0$ moving on a given orbit around the gravastar. In
Ref.~\cite{Gualtieri:2001cm} the solution of Eq.~(\ref{teukolsky}) is
constructed in the general case of elliptic orbits. Eccentricity is expected
to play an important role in extreme mass ratio inspirals
\cite{AmaroSeoane:2007aw,Yunes:2009yz}. However, in the remainder of this
paper we focus on circular inspirals.  This simplifies our study and it is
sufficient to prove our main point: the gravitational radiation from extreme
mass ratio inspirals around horizonless objects is drastically different from
the BH case. We mention in passing that our numerical code is capable of
handling eccentric orbits, and the extension of our study to eccentric
inspirals could be an interesting topic for future research.

We further simplify the problem by using the so-called adiabatic approximation
(i.e. we assume that the radiation reaction timescale is much longer than the
orbital timescale). Under this assumption the trajectory of the particle is
described by the geodesic equations for a mass $m_0$ moving on a circular
orbit of radius $R_0$:
\beq
\bar{\gamma}\equiv\dot t = \frac{E}{1-\frac{2M}{R_0}}, \qquad
\omega_{\rm K}\equiv\frac{d\varphi}{dt}=\frac{\dot{\varphi}}{\bar{\gamma}}\,,
\eeq
where the dot indicates differentiation with respect to proper time.  $E$ is
the energy per unit mass of the particle and $\omega_{\rm K}=\sqrt{M/R_0^3}$
denotes the Keplerian orbital frequency. The source term can be written as
\be
\label{stressenergy}
T_{lm}(\o,r)=\delta(\o-m\o_K)\Bigl[_0S^*_{lm}(\frac{\pi}{2},0)\,_0U_{lm}+
_{-1}S^*_{lm}(\frac{\pi}{2},0)\,_{-1}U_{lm}+\,_{-2}S^*_{lm}(\frac{\pi}{2},0)
\,_{-2}U_{lm}\Bigr]\,,
\ee
where the functions $_{s}U_{lm}$ are explicitly given in
Refs.~\cite{Berti:2002zz,Pons:2001xs}.

The solution of Eq.~(\ref{teukolsky}) satisfying the boundary conditions of
pure outgoing radiation at radial infinity and matching continuously with the
interior solution can be found by the Green's functions technique. The
amplitude of the wave at radial infinity can be shown to be
\cite{Gualtieri:2001cm}
%
\be
\label{ampli0}
A_{l m}(\o)=
-{\frac{1}{W_{l m}(\o)}}~
\int_{R}^{\infty}~ \frac{dr'}{\Delta^2}~\Psi^{~1}_{l m}(\o,r')~ T_{l m}(\o,r')\,,
\ee
where $W_{l m}(\o)$ is the Wronskian of the two independent solutions of the
homogeneous BPT equation
\be
W_{l m}(\o)= \frac{1}{\Delta}\left[
\Psi^{~1}_{l m} \partial_r\Psi^{~0}_{l m}-\Psi^{~0}_{l m} \partial_r\Psi^{~1}_{l m}
\right]\,.\label{wronskian}
\ee
The two solutions $\Psi^{~0}_{l m}$ and $\Psi^{~1}_{l m}$ satisfy different
boundary conditions:
\be
\label{Ro}
\begin{cases}
\dps{
{\cal L}_{BPT}\Psi^{~0}_{l m}(\o,r)=0}\,,
&\cr
\dps{
\Psi^{~0}_{l m}(\o,r\rightarrow\infty) =r^3e^{\ii\o r_*}}\,,
&\cr
\end{cases}
\begin{cases}
\dps{
{\cal L}_{BPT}\Psi^{~1}_{l m}(\o,r)=0}\,, &\cr
\dps{\Psi^{~1}_{l m}(\o,a)= \bar{\Psi}_{l m}(\o,a)}\,,&\cr
\partial_r\dps{\Psi^{~1}_{l m}(\o,a)= \partial_r\bar{\Psi}_{l m}(\o,a)}\,.&\cr
\end{cases}
\ee
Here ${\cal L}_{BPT}$ is the differential operator on the left-hand side of
the BPT equation (\ref{teukolsky}) and $\bar{\Psi}_{l m}(\o,a)$ is the radial
perturbation of the Weyl scalar, constructed according to
Eq.~(\ref{relazione1}) in terms of the perturbed metric functions in the
interior and evaluated at the (exterior) surface of the gravastar.  The
integral in Eq.~(\ref{ampli0}) can be written in terms of $\Psi^{~1}_{l m}$
and its derivatives \cite{Pons:2001xs}.  In Eq.~(\ref{ampli0}) it is
convenient to isolate the contribution of the Dirac $\delta$ function:
\be
\label{ampli}
A_{l m}(\omega)
=m_0\hat A_{l m}(\omega)\delta(\o-m\o_K)\,.
\ee
Then the time-averaged energy-flux
\be
\label{enflux}
\dot E^R\equiv\left<\frac{dE_{GW}}{dt}\right>=\lim_{T\rightarrow\infty}\frac{E_{GW}}{T}
=\lim_{T\rightarrow\infty}\frac{1}{T}\sum_{lm}\int d\omega
\left(\frac{dE_{GW}}{d\omega}\right)_{lm}
\ee
can be written in terms of $\hat A_{lm}(\o)$ as follows:
\be
\label{amplispec}
\dot E^R (m\o_K)
=
\sum_{lm} \f{m_0^2}{4\pi (m \o_K)^2}~
\vert \hat A_{lm}(m\o_K)\vert^2
\equiv \sum_{lm}
\dot{E}^R_{lm}.
\ee
In order to evaluate $\Psi^{~0}_{l m}$ and $\Psi^{~1}_{l m}$, we integrate the
BPT equation by an adaptive Runge-Kutta method. Close to a resonance the
solutions must be computed very accurately, since the Wronskian
(\ref{wronskian}) is the difference between two terms that almost cancel each
other. When required, the tolerance parameter in the adaptive integration
routines is decreased to achieve convergence.  Since the orbital frequency is
related to the orbital velocity $v$ and to the semilatus rectum (which for
circular orbits is simply $p=R_0/M$) by the relations
\be
v=\left( M\o_K\right)^{1/3}=p^{-1/2}\,,
\ee
the energy flux $\dot E^R$ can also be considered as a function of $v$ or $p$.
In the following we shall normalize $\dot E^R$ to the Newtonian quadrupole
energy flux
\be
\label{newtflux}
\dot E^N=\frac{32}{5}\frac{m_0^2}{M^2} v^{10}\,.
\ee
Then the energy flux emitted in gravitational waves normalized to the
Newtonian quadrupole energy flux is given by
\be
P(v) \equiv \frac{\dot{E}^R}{\dot{E}^N}=\sum_{lm}\frac{5}{128\pi} \f{M^2}{(m
\o_K)^2v^{10}}\vert \hat A_{lm}(m\o_K)\vert^2
\,.\label{P(v)}
\ee


The normalized energy flux (\ref{P(v)}) can be computed up to $v\le 1/\sqrt{6}
\simeq 0.408$, which corresponds to the innermost stable circular orbit (ISCO)
at $R_0=6M$. The post-Newtonian expansion of the energy flux $P(v)$ for
particles in circular orbit around Schwarzschild BHs has been studied by
several authors \cite{Poisson:1995vs,Mino:1997bx,Yunes:2008tw}. The
instability of circular orbits with $R_0<6M$ sets an upper bound on the
velocity of the point mass. If the radius of the gravastar is larger than the
ISCO (this typically occurs for $\mu<0.1666$) the upper limit in $v$ will be
smaller.

From the analytical form
of the stress-energy tensor (\ref{stressenergy}) it is easy to see that, for
each assigned $l$, a mode of the star is excited when the orbital frequency
satisfies the resonant condition
\be
\label{cond}
m \omega_{\rm K} =\omega_{\rm QNM}\,,
\ee
where $\omega_{\rm QNM}$ is the QNM frequency. Thus we expect sharp peaks to
appear at the values of $v$ corresponding to the excitation of the gravastar
QNMs for different values of the angular momentum parameter $l$. This offers
an intriguing signature of the absence of event horizons, since the emitted
power for a Schwarzschild BH does not show any peak. In fact one can easily
check that the frequency of the fundamental QNM of a Schwarzschild BH is {\it
  higher} than the critical value $m\omega_{\rm K}$ corresponding to a
particle at the ISCO \cite{Berti:2002zz}. In other words, Schwarzschild QNMs
can only be excited by particles plunging into the BH, while the QNMs of a
gravastar can be excited {\it during the inspiral}. In the following Section
we will compare the power emitted by a circular inspiral around a thin-shell
gravastar to the power emitted by a circular inspiral around a Schwarzschild
BH.

\section{Gravitational flux from gravastars and black holes\label{sec:results}}

Thin-shell gravastar models are specified by two parameters: the gravastar
compactness $\mu=M/a$ and the sound speed parameter $v_s$ that characterizes
the EOS on the shell. Thin-shell gravastars are only one of the several
possible models that can be explored (see
e.g.~\cite{Visser:2003ge,Chirenti:2007mk}) but we expect the qualitative
results of our analysis to apply quite in general. The reason is that the main
difference between gravastars and BHs comes from the different {\it boundary
  conditions} at the ``surface'' replacing the BH event horizon, rather than
from the specific nature of this surface. Furthermore, as discussed below,
peaks in the energy flux are more sensitive to the ``global'' properties of
the gravastar (as determined by the compactness parameter $\mu$) than to the
microphysical model determining the matter distribution on the shell (which in
our simplified case reduces to the specification of a value for $v_s$). Our
numerical study covers the whole range in compactness ($0<\mu<0.5$). We mainly
focused on the most physical range of the EOS parameter ($0<v_s^2<1$) but we
also studied the superluminal case ($v_s^2>1$), and we even allowed for models
with $v_s^2<0$ \cite{Poisson:1995sv}.

The gravitational emission of a Schwarzschild BH perturbed by a particle has
been studied analytically and numerically in great detail for both circular
and eccentric orbits
\cite{Poisson:1993vp,Cutler:1993vq,Cutler:1994pb,Poisson:1995vs}. Our purpose
here is to compare and contrast the energy flux from particles orbiting
Schwarzschild BHs to the energy flux from particles orbiting thin-shell
gravastars. For each value of the gravastar parameters $(\mu,v_s^2)$ we
integrate the perturbations equations (as described in
Section~\ref{sec:perturbations}) for a point-like object of mass $m_0$ moving
on a circular orbit of radius $R_0$ with orbital velocity $v$ and we compute
the energy flux (\ref{P(v)}). Our numerical work uses a modified version of
the BPT code described in Ref.~\cite{Pons:2001xs}. The results obtained by the
BPT formalism were verified using an independent code that integrates the
Zerilli and Regge-Wheeler equations. A slight variant of these codes was used
to compute the flux from a particle orbiting Schwarzschild BHs. The results
are consistent with Refs.~\cite{Poisson:1995vs,Cutler:1993vq} within an
accuracy of about one part in $10^6$ (see Ref.~\cite{Yunes:2008tw} for more
details).

From the results of Paper I, in the Schwarzschild limit $\mu\rightarrow 0.5$
the real part of the QNM frequency tends to zero and to a very good
approximation it is independent of $v_s$. For example, for $\mu=0.49999$ and
$l=2$ we varied $v_s^2$ in the range $[-2,2]$ in steps of $\delta v_s=0.1$ and
we found that the real part of the modes is a constant within a part in $10^6$
($\omega_R=0.235932$), while the imaginary part has tiny variations in the
range between $\omega_I=4.20\times 10^{-7}$ and $\omega_I=4.17\times
10^{-7}$. In order for a QNM to be excited by particles in circular orbits,
the QNM frequency must be small enough to allow for the resonant condition
(\ref{cond}).

\begin{center}
\begin{figure}[ht]
\includegraphics[width=8.8cm,clip=true]{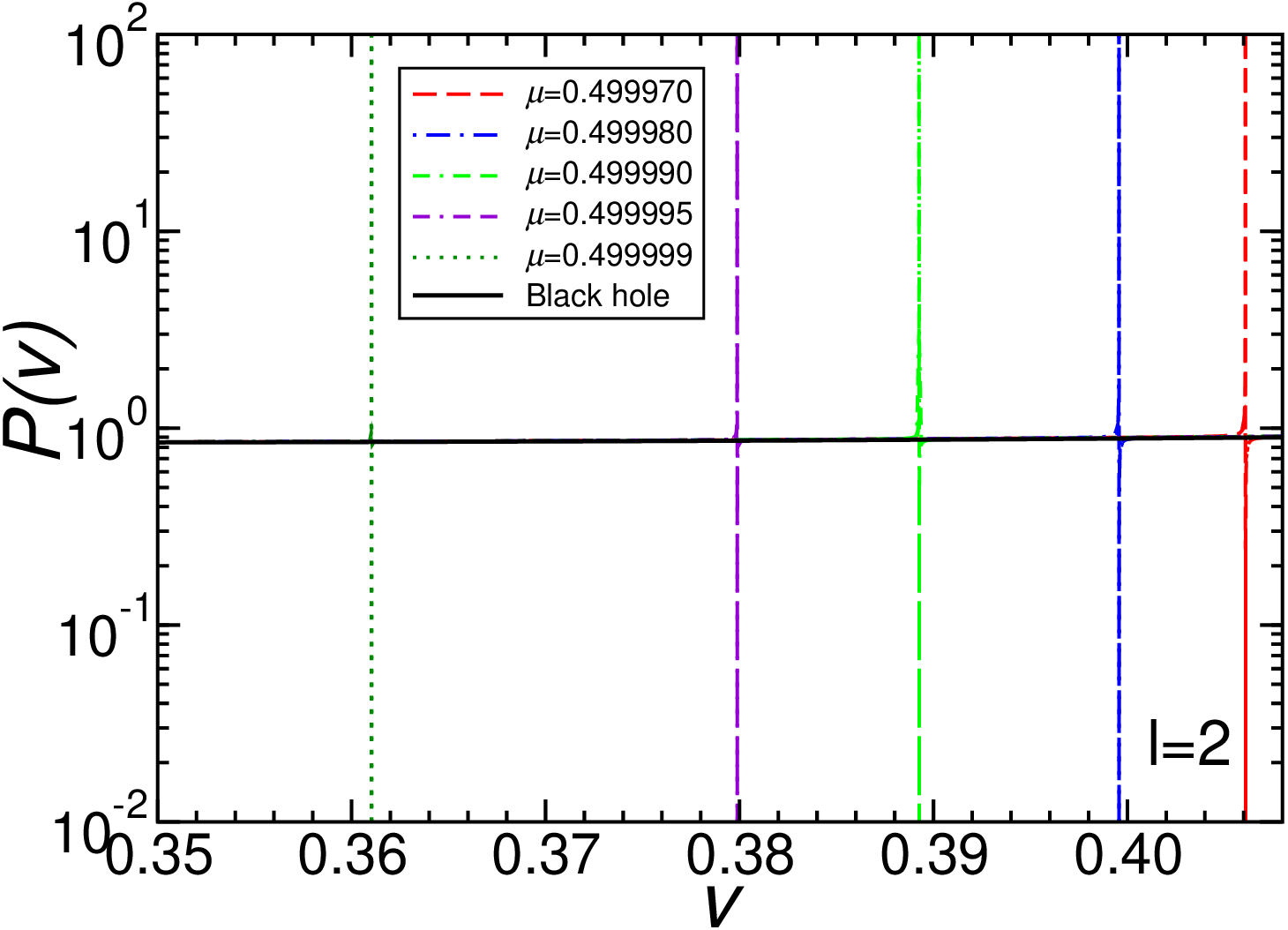}
\caption{Dominant ($l=2$) contribution to the energy flux for very high
  compactness and $v_s^2=0.1$ (but when $\mu\sim0.5$ resonances are almost
  independent on $v_s^2$). From right to left the resonant peaks correspond to
  $\mu=0.49997\,,0.49998\,,0.49999\,,0.499995\,,0.499999$, respectively.}
\label{fig:compact}
\end{figure}
\end{center}

Figure \ref{fig:compact} shows the dominant ($l=2$) contribution to the energy
flux for gravastars with very high compactness.  The frequencies of the lowest
QNMs of a Schwarzschild BH are higher than those of an ultra-compact
gravastar, and cannot be excited by particles in stable circular orbits. For
this reason the power emitted by a BH (on the scale of this plot) is roughly
constant.  Resonance peaks do appear for gravastars, as expected, when
$\omega_{\rm QNM}=2 \omega_{\rm K}$.  Notice that these resonances are
extremely narrow and they would get even narrower for $l>2$. This is because
the imaginary part of the excited modes is extremely small
($2M\omega_I\sim10^{-7},\,10^{-10}$ for $l=2$ and $l=3$ respectively) in the
high-compactness limit $\mu\to 0.5$, leading to a corresponding decrease in
the quality factor of the oscillations. Whether these resonances are actually
detectable is an interesting question for LISA data analysis. The answer
depends on dissipative mechanisms (besides gravitational radiation damping)
that could affect the timescale of the oscillations, especially in the
non-linear regime: see e.g. \cite{Berti:2002ry,Flanagan:2007ix} for
discussions of this problem in the context of neutron star binary detection by
Earth-based GW interferometers.

%
\begin{table}
\centering
\caption{Values of the compactness $\mu$, angular momentum number $l$, QNM
  frequency, orbital velocity $v$ and GW frequency $\nu_{\rm GW}$ of the
  circular orbits which would excite the fundamental QNM of the gravastar for
  the given multipole. The Keplerian frequency is given in mHz and rescaled to
  a gravastar mass $M_6=10^6M_\odot$.}
\vskip 12pt
\begin{tabular}{@{}cccccc@{}}
\hline \hline $\mu$ &$l$ &$M\omega_{\rm QNM}$ &$v$ &$(M_6/M) \nu_{\rm GW}$
(mHz) \\ 

\hline 0.49997  &2 &0.1339 &0.4061 &4.328\\ 
                &3 &0.1508 &0.3691 &4.873\\
\hline 0.49998  &2 &0.1276 &0.3996 &4.123\\
                &3 &0.1429 &0.3625 &4.616\\
\hline 0.49999  &2 &0.1180 &0.3893 &3.812\\
                &3 &0.1310 &0.3521 &4.232\\
\hline 0.499995 &2 &0.1096 &0.3799 &3.543\\
\hline 0.499999 &2 &0.0941 &0.3610 &3.041\\
\hline
\hline
\end{tabular}
\label{table1}
\end{table}
%

Quite interestingly, gravastars that ``try harder'' to look like a BH (in the
sense that their shell is closer to the Schwarzschild event horizon) are those
for which the peak in the energy flux appears for smaller values of
$\mu$. Table \ref{table1} lists the expected excited modes for different
values of $\mu$ corresponding to ultra-compact gravastars.

One may worry that the resonance will eventually get out of the LISA band for
gravastars having $\mu$ {\it extremely} close to the Schwarzschild value. The
following naive argument suggests that this is not the case. The ``thick shell
gravastar'' model by Mazur and Mottola predicts a microscopic but finite shell
thickness $\ell \sim \sqrt{L_{\rm Pl} r_{\rm S}}\simeq 3\times
10^{-14}(M/M_\odot)^{1/2}$~cm, where $L_{\rm Pl}$ is the Planck scale and
$r_{\rm S}$ is the Schwarzschild radius, so that the energy density and
pressure in the shell are far below Planckian and the geometry can still be
described reliably by Einstein's equations \cite{Mazur:2004fk}. Our simplified
model does not allow for a finite thickness of the shell, and a microscopic
model of finite shells is required for a careful analysis of this problem.
However, for the sake of argument, let us consider $\epsilon=1/2-\mu\to 0$ as
a ``thickness parameter'' describing how far the gravastar shell can be
relative to the BH horizon. A power-law fit of the QNMs of a thin-shell
gravastar in the limit $\epsilon\rightarrow0$ yields $f_{\rm GW}\sim
3.828\times (\epsilon\times 10^{-5})^{0.1073}$.
%
The lower frequency sensitivity limit for LISA is dictated by acceleration
noise. Assuming lower frequency cutoffs of $f_{\rm low}=10^{-5}, 3\times
10^{-5}, 10^{-4}$, we find that the peaks will sweep out of the LISA band when
$\epsilon=9.6\times 10^{-48}, 2.7\times 10^{-43}, 2.0\times 10^{-38}$,
respectively.
This estimate of the ``minimum measurable deviation from a BH'' is admittedly
very sensitive to the fitting function we use and it may change when one
considers thick shell gravastars, but it suggests that LISA has the potential
to reveal solid surfaces replacing horizons even when these solid surfaces are
very close to the location of the Schwarzschild horizon.
%

%
\begin{center}
\begin{figure}[ht]
\begin{tabular}{cc}
\includegraphics[width=8.8cm,clip=true]{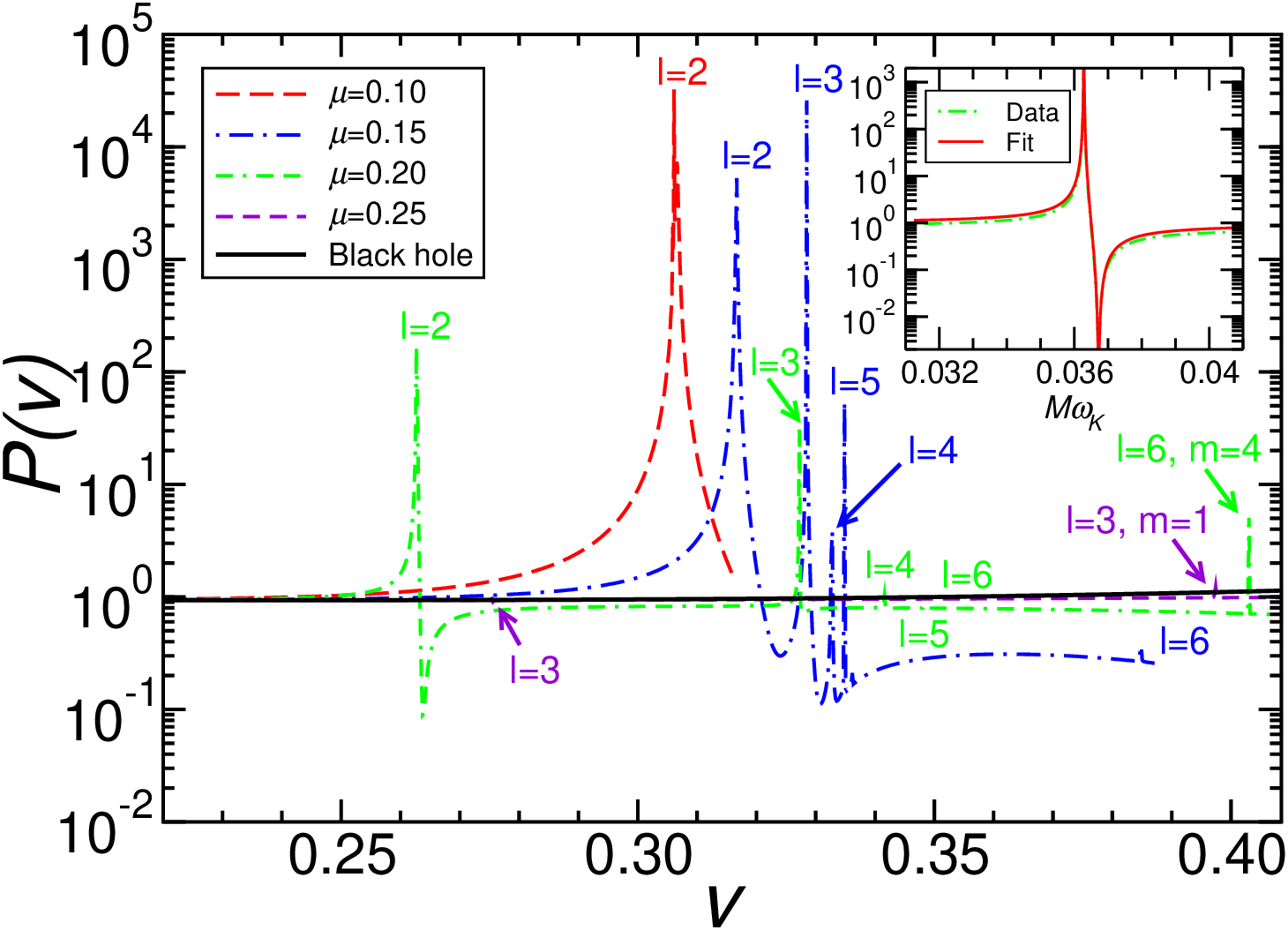} &
\includegraphics[width=8.8cm,clip=true]{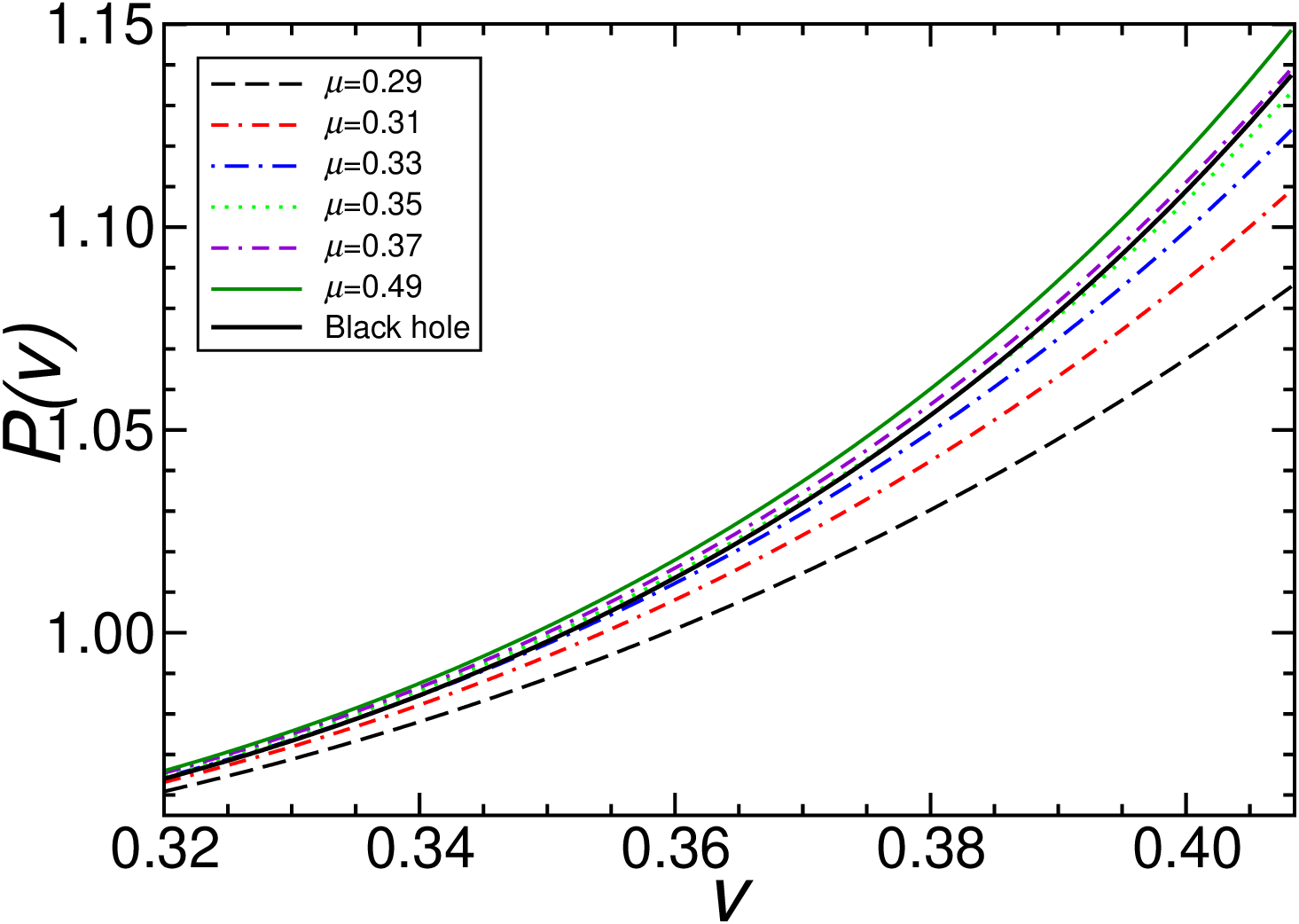}\\
\end{tabular}
\caption{Left: The energy flux (summed up to $l=6$) of GWs emitted by a small
  mass orbiting thin-shell gravastars with $v_s^2=0.1$ and different values of
  $\mu$ (plotted as a function of the particle orbital velocity $v$) is
  compared with the flux for a Schwarzschild BH. All peaks (with the exception
  of the last two peaks on the right) are due to the excitation of QNMs with
  $l=m$.  Right: same for $v_s^2=0.1$ and selected values of $\mu\in [0.29,
    0.49]$. No QNMs are excited in this range.}
\label{fig:noncompact}
\end{figure}
\end{center}

A relevant question is whether massive horizonless objects which are compact
by the standard of (say) main sequence stars, but ``only'' as compact as
neutron stars, can leave a signature on the gravitational signal emitted by
small, inspiralling compact objects. In Fig.~\ref{fig:noncompact} we plot the
normalized energy flux $P(v)$ as a function of the orbital velocity for
gravastar models with $v_s^2=0.1$ and compactness in the range $0.1\lesssim
\mu\lesssim 0.49$, as well as for a Schwarzschild BH.  The total flux was
computed by adding all multipoles ($|m|\leq l$) and by truncating the
multipolar expansion at $l=6$. As discussed in
Refs.~\cite{Poisson:1993vp,Gualtieri:2001cm,Berti:2002zz,Yunes:2008tw}, a
multipole of order $l$ contributes to the total power as a correction of order
$p^{2-l}$. Roughly speaking, a truncation at $l=6$ produces a relative error
(in the non-resonant regime) of order $p^{-5}=v^{10}$ (but see
\cite{Yunes:2008tw} for a more careful discussion of the convergence
properties of the post-Newtonian series). When $\mu\gtrsim 0.166$ the ISCO is
located outside the gravastar and we plot the energy flux up to the ISCO
velocity $v_{\rm ISCO}\simeq 0.408$ (corresponding to $R_0=6 M$). For less
compact gravastars, plots of the energy flux are truncated at the velocity
corresponding to the location of the shell.

\begin{center}
\begin{figure}[ht]
\begin{tabular}{cc}
\includegraphics[width=8.8cm,clip=true]{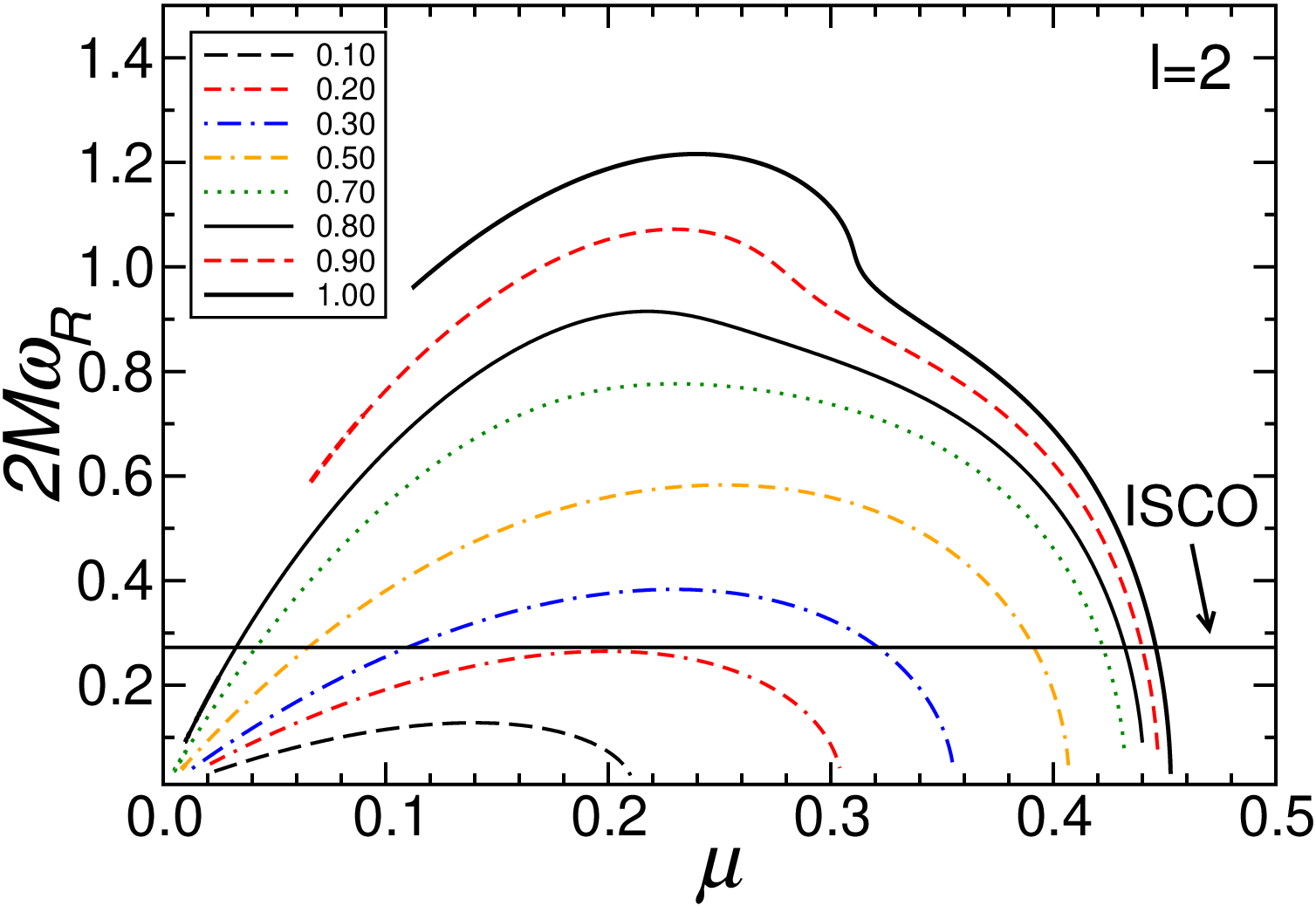}&
\includegraphics[width=8.8cm,clip=true]{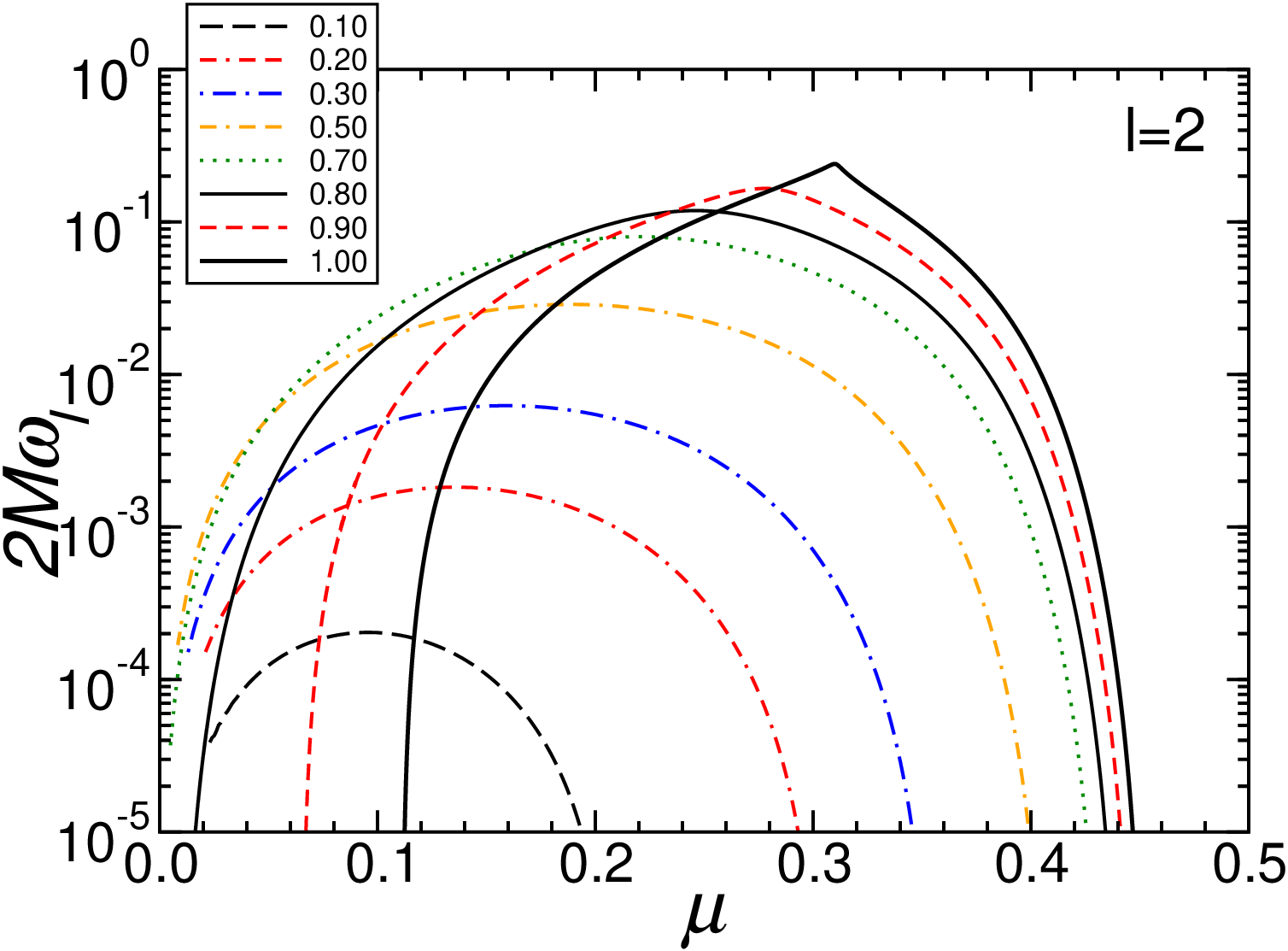}\\
\end{tabular}
\caption{Real part (left) and imaginary part (right) of gravastar QNMs with
  $l=2$ as a function of compactness for several fixed values of $v_s^2$ (as
  indicated in the legend). For clarity in illustrating the ``selection
  rules'' that determine QNM excitation during inspiral we only show the
  weakly damped part of the QNM spectrum (compare Fig.~2 and Fig.~6 in Paper
  I). For $v_s^2>0.8$ the real part of the frequency is plotted down to the
  critical minimum compactness at which the imaginary part crosses zero within
  our numerical accuracy.  the horizontal line at $2M\omega_R\simeq 0.2722$
  corresponds to twice the orbital frequency of a particle in circular orbit
  at the ISCO: only QNMs below this line can be excited during a
  quasi-circular inspiral.}
\label{fig:modesreim}
\end{figure}
\end{center}
\begin{center}
\begin{figure}[ht]
\begin{tabular}{cc}
\includegraphics[width=8.8cm,clip=true]{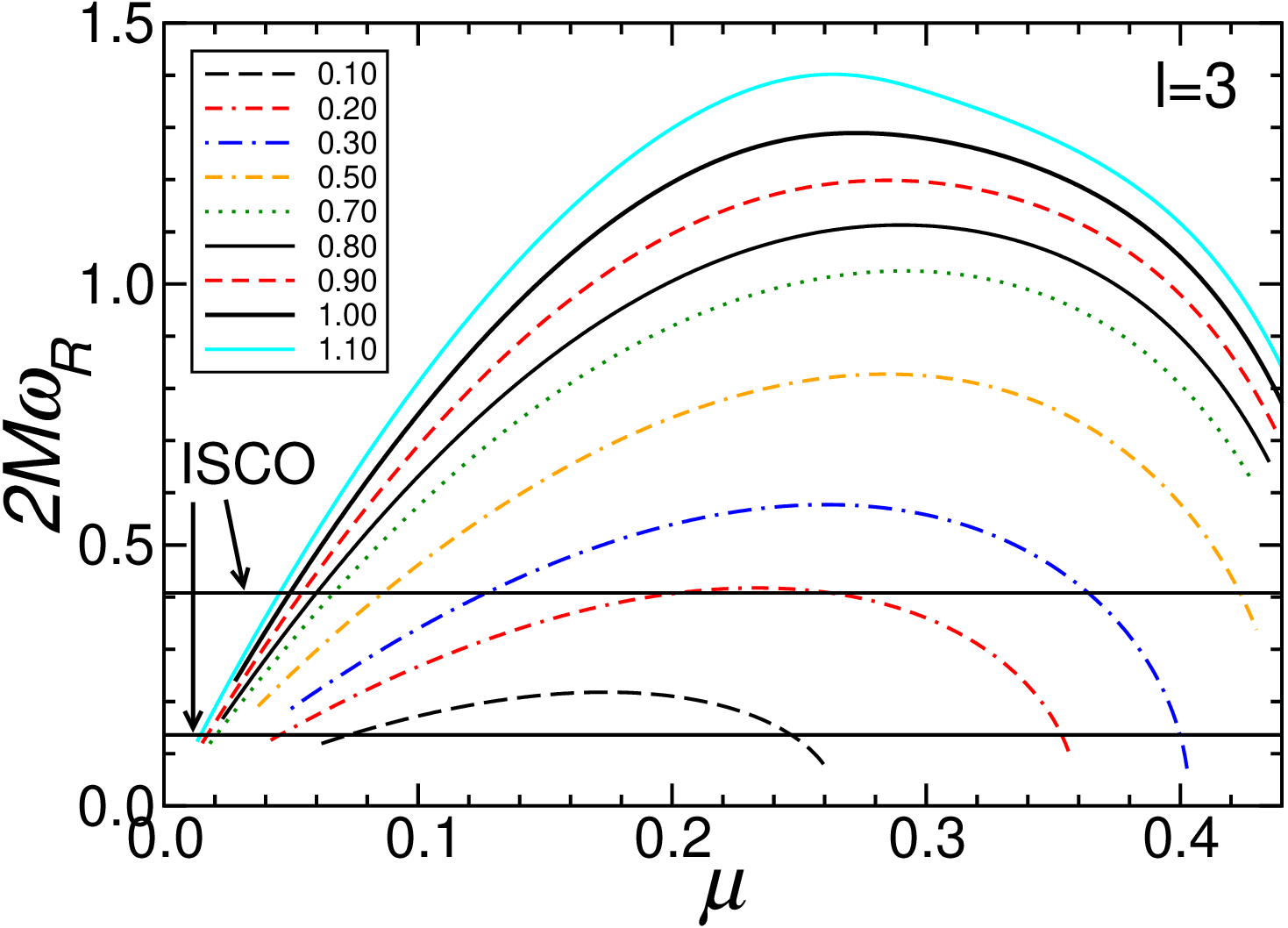}&
\includegraphics[width=8.8cm,clip=true]{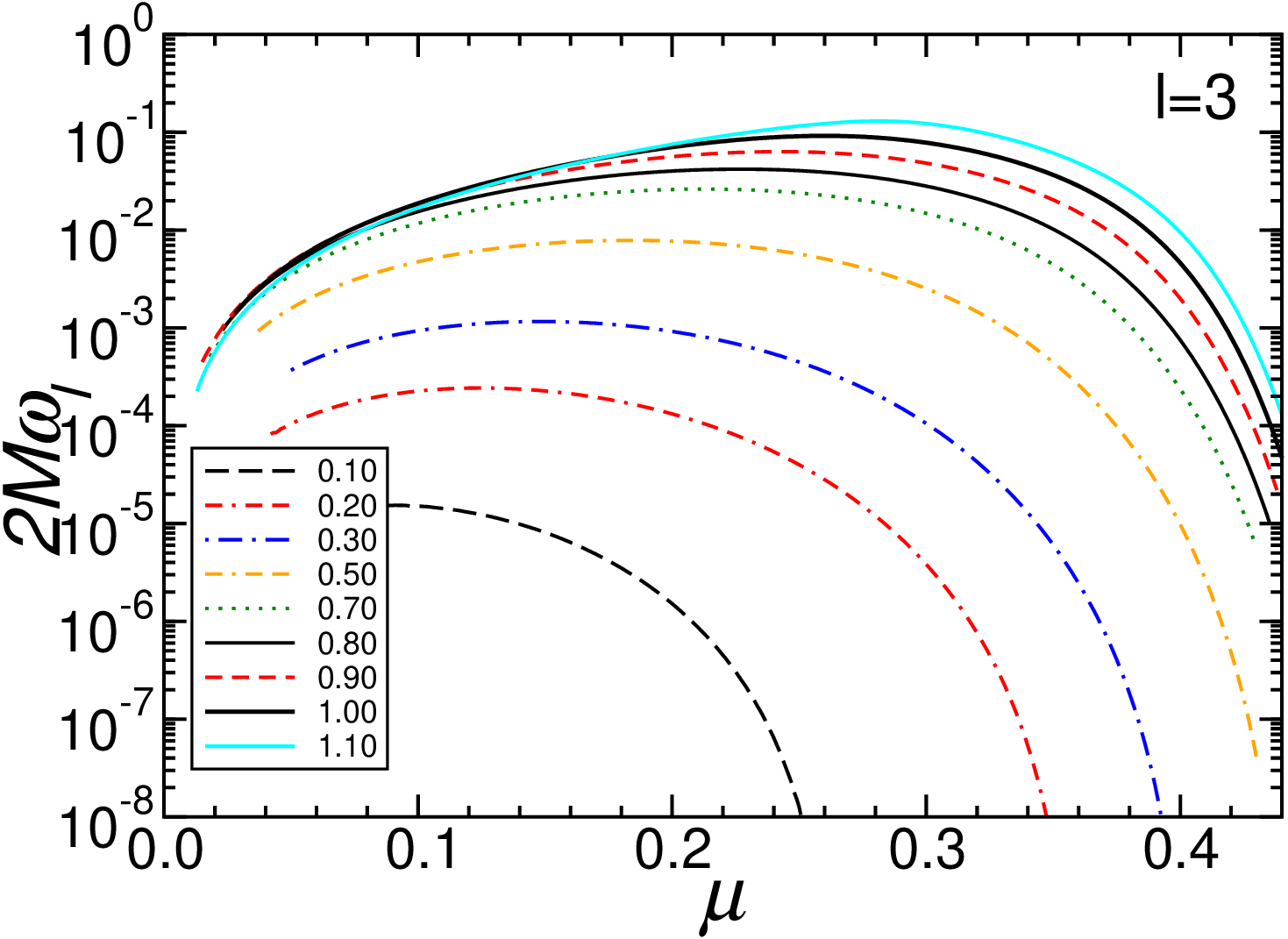}
\end{tabular}
\caption{Real part (left) and imaginary part (right) of gravastar QNMs with
  $l=3$ as a function of compactness for several fixed values of $v_s^2$ (as
  indicated in the legend). For clarity in illustrating the ``selection
  rules'' that determine QNM excitation during inspiral we only show the
  weakly damped part of the QNM spectrum. In the left panel, the horizontal
  lines at $2M\omega_R\simeq 0.1361$ ($2M\omega_R\simeq 0.4082$) correspond to
  the orbital frequency (or three times the orbital frequency) of a particle
  in circular orbit at the ISCO. Perturbations with $l=3, m=1$ can excite the
  QNMs below the first line, while perturbations with $l=m=3$ can excite QNMs
  below the second line.}
\label{fig:modesreim_l3}
\end{figure}
\end{center}

The complex structure of the spectrum for values of $\mu$ smaller than about
$0.2$ is best understood by considering the real and imaginary parts of the
weakly damped QNM frequencies of a gravastar (see
Fig.~\ref{fig:modesreim}). For clarity in Fig.~\ref{fig:modesreim} we only
plot weakly damped QNMs, but our general arguments apply also to the second,
``ordinary'' family of QNMs (cf.~Paper I). In particular, from Fig.~2 and
Fig.~6 in Paper I it should be clear that QNMs will be excited for low values
of the compactness {\it and} when $\mu$ is very close to the BH value
$\mu=1/2$. Besides these ``ultracompact'' modes, only QNMs whose real part
lies {\it below} the horizontal line in the left panel (corresponding to twice
the ISCO orbital frequency for a particle in circular orbit) can be excited.

Fig.~\ref{fig:modesreim} clarifies that the range of $\mu$ over which QNMs can
be excited depends on $v_s$.  For $v_s^2=0.1$ (the case considered to produce
the energy fluxes of Fig.~\ref{fig:noncompact}) QNM frequencies that can be
excited by resonant inspirals only exist for $\mu\lesssim 0.21$ (left panel of
Fig.~\ref{fig:noncompact}) or for $\mu\gtrsim 0.49997$, i.e. when the thin
shell is extremely close to the location of the BH horizon
(Fig.~\ref{fig:compact}). The real part of the corresponding QNM frequency has
a local maximum at $\mu\approx 0.15$. Correspondingly, the $l=2$ QNM peak
visible in the energy flux of Fig.~\ref{fig:noncompact} occurs later in the
inspiral for the $\mu=0.15$ model than it does for the $\mu=0.10$ and
$\mu=0.20$ models.

In Fig.~\ref{fig:noncompact} the $l=2$ and $l=3$ peaks for $\mu=0.20$ are well
separated in frequency and an ``antiresonance'' is visible to the right of the
$l=2$ resonance. The nature of this antiresonance can be explained by a simple
harmonic oscillator model \cite{Pons:2001xs}. In the inset of the left panel
of Fig.~\ref{fig:noncompact} we plot {\it both} the resonance and
antiresonance as functions of the Keplerian orbital frequency of the particle
$M\omega_{\rm K}$ for $\mu=0.2$ and $l=2$ (dashed green line). A fit using the
simple harmonic oscillator model of Ref.~\cite{Pons:2001xs} (red line)
reproduces the qualitative features of both resonance and antiresonance: in
this specific case the fit gives $2M\omega_R\sim0.07257$ and
$2M\omega_I\sim2\times10^{-6}$, while QNM calculations using the resonance
method yield $2M\omega_R\sim0.07257$ and $2M\omega_I\sim4\times10^{-6}$.

Modes with $l>2$ are typically harder to excite because of their higher
frequencies and lower quality factors. However, because of the complex
``selection rules'' illustrated in Fig.~\ref{fig:modesreim} for $l=2$,
sometimes only resonances with $l>2$ will be visible. When $v_s^2=0.1$ and
$\mu>0.21$ only modes with $l>2$ can be excited, and only narrow $l=3$
resonances can be seen in Fig.~\ref{fig:noncompact} when the compactness
$\mu=0.25$ (cf. Fig.~\ref{fig:modesreim_l3}).

When $l=2$ the imaginary part of one QNM with $v_s^2=0.1$ crosses zero within
our numerical accuracy at the ``critical'' compactness $\mu\simeq 0.21$,
possibly signaling a (marginal) nonradial instability of the gravastar, and no
QNMs can be excited for $0.21\leq \mu \leq 0.49997$. In this compactness range
the energy flux emitted by either the gravastar or the BH is mostly due to the
orbital motion and it only depends on the compactness of the central object.
The right panel of Fig.~\ref{fig:noncompact} shows that the flux emitted by a
gravastar approaches the BH flux ``from below'' as the compactness
increases. For $\mu\simeq 0.35$ the gravastar flux is almost indistinguishable
from the BH flux and for $\mu>0.35$ a gravastar radiates slightly more
than a BH. This is due to the fact that the emitted power ``feels'' the
contribution of resonances, which in this case correspond to orbits smaller
than the ISCO but do nevertheless contribute to increase the slope of the
curve. A similar trend can be seen in neutron star calculations in regions of
the parameter space where the contribution from resonances is negligible
\cite{Pons:2001xs}.

If gravastars or other horizonless objects have astrophysical reality, the
presence or absence of resonant peaks in the GW flux can provide interesting
information on the microscopic properties of the physical surface replacing
the event horizon.  Suppose for example that we can estimate the compactness
of a massive object by independent means (e.g. by electromagnetic
observations). Even within our simple thin-shell model, the range in frequency
where resonances in the GW emission from EMRIs are allowed changes with
$v_s^2$. For example, if $v_s^2=0.1$ resonances can exist when $\mu\lesssim
0.21$ or $\mu\sim 0.5$, but if $v_s^2=0.3$ they can exist when $\mu \lesssim
0.1$, $0.33\lesssim \mu\lesssim 0.36$ and $\mu\sim 0.5$.  Similar results also
hold when $l=3$, as shown in Fig.~\ref{fig:modesreim_l3}. For example, if
$v_s^2=0.1$ resonances can exist when $\mu\lesssim 0.27$ or $\mu\sim 0.5$, but
if $v_s^2=0.3$ they can exist when $\mu \lesssim 0.2$, $0.37\lesssim
\mu\lesssim 0.41$ and $\mu\sim 0.5$. So, in general, the range of $\mu$ where
QNM frequencies can be excited by a circular inspiral depend on the value of
$l$. In the Schwarzschild limit ($\mu\sim0.5$) QNM frequencies are excited for
any $l$, but higher--$l$ modes have a tiny imaginary part
($2M\omega_I\sim10^{-10}$ for $l=3$) and they are more difficult to detect
than the dominant ($l=2$) modes.

If an EMRI is detected, the existence of these selection rules (in the form of
compactness regions where resonances can or cannot exist) in principle allows
for null tests of the existence of an event horizon for objects of the given
compactness. Similar arguments can presumably be made for more complex (or
contrived) gravastar models.

%
%
%

\section{Conclusions}

We have shown that gravitational radiation from EMRIs can be used to tell the
presence or absence of an event horizon in a compact, massive object. More
specifically, we have shown that the resonant excitation of the oscillation
modes of a gravastar in the LISA band is a potentially observable signature of
the surface replacing the event horizon.  For thin-shell gravastar models
there is a range of compactness (e.g., $\mu\lesssim 0.21$ and $\mu
\gtrsim0.49997$ for $v_s^2=0.1$) where this resonant scattering can occur. 

More detailed data analysis studies (possibly including refined microphysical
models of this ``solid surface'') are necessary to determine the detectability
of resonant peaks, especially for ultra-compact gravastars. The extension of
our results to rotating gravastar models presents a challenge because of the
difficulties in finding plausible rotating gravastar solutions and because of
the ergoregion instability that affects some rotating gravastar models
\cite{Cardoso:2007az,Cardoso:2008kj,Chirenti:2008pf}.

\section*{Acknowledgements}
P.P. thanks the Department of Physics, University of Rome ``La Sapienza'' for the kind hospitality. 
E.B.'s research was supported by NSF grant PHY-0900735. V.C. was supported by
a ``Ci\^encia 2007'' research contract and by Funda\c c\~ao Calouste
Gulbenkian through a short-term scholarship. Y.C. was supported by NSF grants
PHY-0653653 and PHY-0601459, and the David and Barbara Groce Start-up Fund at
Caltech. This work was partially supported by FCT - Portugal through projects
PTDC/FIS/64175/2006, PTDC/FIS/098025/2008, PTDC/FIS/098032/2008
PTDC/CTE-AST/098034/2008, CERN/FP/109290/2009. The authors thankfully
acknowledge the computer resources, technical expertise and assistance
provided by the Barcelona Supercomputing Centre---Centro Nacional de
Supercomputaci\'on.
\bibliography{gravastar_EMRIs}

\begin{thebibliography}{37}
\expandafter\ifx\csname natexlab\endcsname\relax\def\natexlab#1{#1}\fi
\expandafter\ifx\csname bibnamefont\endcsname\relax
  \def\bibnamefont#1{#1}\fi
\expandafter\ifx\csname bibfnamefont\endcsname\relax
  \def\bibfnamefont#1{#1}\fi
\expandafter\ifx\csname citenamefont\endcsname\relax
  \def\citenamefont#1{#1}\fi
\expandafter\ifx\csname url\endcsname\relax
  \def\url#1{\texttt{#1}}\fi
\expandafter\ifx\csname urlprefix\endcsname\relax\def\urlprefix{URL }\fi
\providecommand{\bibinfo}[2]{#2}
\providecommand{\eprint}[2][]{\url{#2}}

\bibitem[{\citenamefont{Narayan}(2005)}]{Narayan:2005ie}
\bibinfo{author}{\bibfnamefont{R.}~\bibnamefont{Narayan}},
  \bibinfo{journal}{New J. Phys.} \textbf{\bibinfo{volume}{7}},
  \bibinfo{pages}{199} (\bibinfo{year}{2005}), \eprint{gr-qc/0506078}.

\bibitem[{\citenamefont{Psaltis}(2008)}]{Psaltis:2008bb}
\bibinfo{author}{\bibfnamefont{D.}~\bibnamefont{Psaltis}}
  (\bibinfo{year}{2008}), \eprint{0806.1531}.

\bibitem[{\citenamefont{Visser}(2009)}]{Visser:2009xp}
\bibinfo{author}{\bibfnamefont{M.}~\bibnamefont{Visser}}
  (\bibinfo{year}{2009}), \eprint{0901.4365}.

\bibitem[{\citenamefont{Abramowicz et~al.}(2002)\citenamefont{Abramowicz,
  Kluzniak, and Lasota}}]{Abramowicz:2002vt}
\bibinfo{author}{\bibfnamefont{M.~A.} \bibnamefont{Abramowicz}},
  \bibinfo{author}{\bibfnamefont{W.}~\bibnamefont{Kluzniak}}, \bibnamefont{and}
  \bibinfo{author}{\bibfnamefont{J.-P.} \bibnamefont{Lasota}},
  \bibinfo{journal}{Astron. Astrophys.} \textbf{\bibinfo{volume}{396}},
  \bibinfo{pages}{L31} (\bibinfo{year}{2002}), \eprint{astro-ph/0207270}.

\bibitem[{\citenamefont{Sathyaprakash and Schutz}(2009)}]{Sathyaprakash:2009xs}
\bibinfo{author}{\bibfnamefont{B.~S.} \bibnamefont{Sathyaprakash}}
  \bibnamefont{and} \bibinfo{author}{\bibfnamefont{B.~F.}
  \bibnamefont{Schutz}}, \bibinfo{journal}{Living Rev. Rel.}
  \textbf{\bibinfo{volume}{12}}, \bibinfo{pages}{2} (\bibinfo{year}{2009}),
  \eprint{0903.0338}.

\bibitem[{\citenamefont{Kokkotas and Schmidt}(1999)}]{Kokkotas:1999bd}
\bibinfo{author}{\bibfnamefont{K.~D.} \bibnamefont{Kokkotas}} \bibnamefont{and}
  \bibinfo{author}{\bibfnamefont{B.~G.} \bibnamefont{Schmidt}},
  \bibinfo{journal}{Living Rev. Rel.} \textbf{\bibinfo{volume}{2}},
  \bibinfo{pages}{2} (\bibinfo{year}{1999}), \eprint{gr-qc/9909058}.

\bibitem[{\citenamefont{Berti et~al.}(2009)\citenamefont{Berti, Cardoso, and
  Starinets}}]{Berti:2009kk}
\bibinfo{author}{\bibfnamefont{E.}~\bibnamefont{Berti}},
  \bibinfo{author}{\bibfnamefont{V.}~\bibnamefont{Cardoso}}, \bibnamefont{and}
  \bibinfo{author}{\bibfnamefont{A.~O.} \bibnamefont{Starinets}},
  \bibinfo{journal}{Class. Quant. Grav.} \textbf{\bibinfo{volume}{26}},
  \bibinfo{pages}{163001} (\bibinfo{year}{2009}), \eprint{0905.2975}.

\bibitem[{\citenamefont{Pani et~al.}(2009)\citenamefont{Pani, Berti, Cardoso,
  Chen, and Norte}}]{Pani:2009ss}
\bibinfo{author}{\bibfnamefont{P.}~\bibnamefont{Pani}},
  \bibinfo{author}{\bibfnamefont{E.}~\bibnamefont{Berti}},
  \bibinfo{author}{\bibfnamefont{V.}~\bibnamefont{Cardoso}},
  \bibinfo{author}{\bibfnamefont{Y.}~\bibnamefont{Chen}}, \bibnamefont{and}
  \bibinfo{author}{\bibfnamefont{R.}~\bibnamefont{Norte}}
  (\bibinfo{year}{2009}), \eprint{0909.0287}.

\bibitem[{\citenamefont{Fiziev}(2006)}]{Fiziev:2005ki}
\bibinfo{author}{\bibfnamefont{P.~P.} \bibnamefont{Fiziev}},
  \bibinfo{journal}{Class. Quant. Grav.} \textbf{\bibinfo{volume}{23}},
  \bibinfo{pages}{2447} (\bibinfo{year}{2006}), \eprint{gr-qc/0509123}.

\bibitem[{\citenamefont{Chirenti and Rezzolla}(2007)}]{Chirenti:2007mk}
\bibinfo{author}{\bibfnamefont{C.~B. M.~H.} \bibnamefont{Chirenti}}
  \bibnamefont{and} \bibinfo{author}{\bibfnamefont{L.}~\bibnamefont{Rezzolla}},
  \bibinfo{journal}{Class. Quant. Grav.} \textbf{\bibinfo{volume}{24}},
  \bibinfo{pages}{4191} (\bibinfo{year}{2007}), \eprint{0706.1513}.

\bibitem[{\citenamefont{Ryan}(1995)}]{Ryan:1995wh}
\bibinfo{author}{\bibfnamefont{F.~D.} \bibnamefont{Ryan}},
  \bibinfo{journal}{Phys. Rev.} \textbf{\bibinfo{volume}{D52}},
  \bibinfo{pages}{5707} (\bibinfo{year}{1995}).

\bibitem[{\citenamefont{Ryan}(1997)}]{Ryan:1997hg}
\bibinfo{author}{\bibfnamefont{F.~D.} \bibnamefont{Ryan}},
  \bibinfo{journal}{Phys. Rev.} \textbf{\bibinfo{volume}{D56}},
  \bibinfo{pages}{1845} (\bibinfo{year}{1997}).

\bibitem[{\citenamefont{Li and Lovelace}(2008)}]{Li:2007qu}
\bibinfo{author}{\bibfnamefont{C.}~\bibnamefont{Li}} \bibnamefont{and}
  \bibinfo{author}{\bibfnamefont{G.}~\bibnamefont{Lovelace}},
  \bibinfo{journal}{Phys. Rev.} \textbf{\bibinfo{volume}{D77}},
  \bibinfo{pages}{064022} (\bibinfo{year}{2008}), \eprint{gr-qc/0702146}.

\bibitem[{\citenamefont{Kojima}(1987)}]{Kojima:1987tk}
\bibinfo{author}{\bibfnamefont{Y.}~\bibnamefont{Kojima}},
  \bibinfo{journal}{Prog. Theor. Phys.} \textbf{\bibinfo{volume}{77}},
  \bibinfo{pages}{297} (\bibinfo{year}{1987}).

\bibitem[{\citenamefont{Gualtieri et~al.}(2001)\citenamefont{Gualtieri, Berti,
  Pons, Miniutti, and Ferrari}}]{Gualtieri:2001cm}
\bibinfo{author}{\bibfnamefont{L.}~\bibnamefont{Gualtieri}},
  \bibinfo{author}{\bibfnamefont{E.}~\bibnamefont{Berti}},
  \bibinfo{author}{\bibfnamefont{J.~A.} \bibnamefont{Pons}},
  \bibinfo{author}{\bibfnamefont{G.}~\bibnamefont{Miniutti}}, \bibnamefont{and}
  \bibinfo{author}{\bibfnamefont{V.}~\bibnamefont{Ferrari}},
  \bibinfo{journal}{Phys. Rev.} \textbf{\bibinfo{volume}{D64}},
  \bibinfo{pages}{104007} (\bibinfo{year}{2001}), \eprint{gr-qc/0107046}.

\bibitem[{\citenamefont{Pons et~al.}(2002)\citenamefont{Pons, Berti, Gualtieri,
  Miniutti, and Ferrari}}]{Pons:2001xs}
\bibinfo{author}{\bibfnamefont{J.~A.} \bibnamefont{Pons}},
  \bibinfo{author}{\bibfnamefont{E.}~\bibnamefont{Berti}},
  \bibinfo{author}{\bibfnamefont{L.}~\bibnamefont{Gualtieri}},
  \bibinfo{author}{\bibfnamefont{G.}~\bibnamefont{Miniutti}}, \bibnamefont{and}
  \bibinfo{author}{\bibfnamefont{V.}~\bibnamefont{Ferrari}},
  \bibinfo{journal}{Phys. Rev.} \textbf{\bibinfo{volume}{D65}},
  \bibinfo{pages}{104021} (\bibinfo{year}{2002}), \eprint{gr-qc/0111104}.

\bibitem[{\citenamefont{Berti et~al.}(2002)\citenamefont{Berti, Pons, Miniutti,
  Gualtieri, and Ferrari}}]{Berti:2002ry}
\bibinfo{author}{\bibfnamefont{E.}~\bibnamefont{Berti}},
  \bibinfo{author}{\bibfnamefont{J.~A.} \bibnamefont{Pons}},
  \bibinfo{author}{\bibfnamefont{G.}~\bibnamefont{Miniutti}},
  \bibinfo{author}{\bibfnamefont{L.}~\bibnamefont{Gualtieri}},
  \bibnamefont{and} \bibinfo{author}{\bibfnamefont{V.}~\bibnamefont{Ferrari}},
  \bibinfo{journal}{Phys. Rev.} \textbf{\bibinfo{volume}{D66}},
  \bibinfo{pages}{064013} (\bibinfo{year}{2002}), \eprint{gr-qc/0208011}.

\bibitem[{\citenamefont{Kesden et~al.}(2005)\citenamefont{Kesden, Gair, and
  Kamionkowski}}]{Kesden:2004qx}
\bibinfo{author}{\bibfnamefont{M.}~\bibnamefont{Kesden}},
  \bibinfo{author}{\bibfnamefont{J.}~\bibnamefont{Gair}}, \bibnamefont{and}
  \bibinfo{author}{\bibfnamefont{M.}~\bibnamefont{Kamionkowski}},
  \bibinfo{journal}{Phys. Rev.} \textbf{\bibinfo{volume}{D71}},
  \bibinfo{pages}{044015} (\bibinfo{year}{2005}), \eprint{astro-ph/0411478}.

\bibitem[{\citenamefont{Cutler et~al.}(1993)\citenamefont{Cutler, Poisson,
  Sussman, and Finn}}]{Cutler:1993vq}
\bibinfo{author}{\bibfnamefont{C.}~\bibnamefont{Cutler}},
  \bibinfo{author}{\bibfnamefont{E.}~\bibnamefont{Poisson}},
  \bibinfo{author}{\bibfnamefont{G.~J.} \bibnamefont{Sussman}},
  \bibnamefont{and} \bibinfo{author}{\bibfnamefont{L.~S.} \bibnamefont{Finn}},
  \bibinfo{journal}{Phys. Rev.} \textbf{\bibinfo{volume}{D47}},
  \bibinfo{pages}{1511} (\bibinfo{year}{1993}).

\bibitem[{\citenamefont{Cutler et~al.}(1994)\citenamefont{Cutler, Kennefick,
  and Poisson}}]{Cutler:1994pb}
\bibinfo{author}{\bibfnamefont{C.}~\bibnamefont{Cutler}},
  \bibinfo{author}{\bibfnamefont{D.}~\bibnamefont{Kennefick}},
  \bibnamefont{and} \bibinfo{author}{\bibfnamefont{E.}~\bibnamefont{Poisson}},
  \bibinfo{journal}{Phys. Rev.} \textbf{\bibinfo{volume}{D50}},
  \bibinfo{pages}{3816} (\bibinfo{year}{1994}).

\bibitem[{\citenamefont{Poisson}(1995)}]{Poisson:1995vs}
\bibinfo{author}{\bibfnamefont{E.}~\bibnamefont{Poisson}},
  \bibinfo{journal}{Phys. Rev.} \textbf{\bibinfo{volume}{D52}},
  \bibinfo{pages}{5719} (\bibinfo{year}{1995}), \eprint{gr-qc/9505030}.

\bibitem[{\citenamefont{Visser and Wiltshire}(2004)}]{Visser:2003ge}
\bibinfo{author}{\bibfnamefont{M.}~\bibnamefont{Visser}} \bibnamefont{and}
  \bibinfo{author}{\bibfnamefont{D.~L.} \bibnamefont{Wiltshire}},
  \bibinfo{journal}{Class. Quant. Grav.} \textbf{\bibinfo{volume}{21}},
  \bibinfo{pages}{1135} (\bibinfo{year}{2004}), \eprint{gr-qc/0310107}.

\bibitem[{\citenamefont{Abramowitz and Stegun}(1972)}]{Abramowitz:1970as}
\bibinfo{author}{\bibfnamefont{M.}~\bibnamefont{Abramowitz}} \bibnamefont{and}
  \bibinfo{author}{\bibfnamefont{I.~A.} \bibnamefont{Stegun}},
  \emph{\bibinfo{title}{Handbook of Mathematical Functions with Formulas,
  Graphs, and Mathematical Tables}} (\bibinfo{publisher}{Dover},
  \bibinfo{address}{New York}, \bibinfo{year}{1972}).

\bibitem[{\citenamefont{Berti}(2002)}]{Berti:2002zz}
\bibinfo{author}{\bibfnamefont{E.}~\bibnamefont{Berti}} (\bibinfo{year}{2002}),
  \bibinfo{note}{{Ph.D. thesis (unpublished)}}.

\bibitem[{\citenamefont{Bardeen and Press}(1973)}]{Bardeen:1973xb}
\bibinfo{author}{\bibfnamefont{J.~M.} \bibnamefont{Bardeen}} \bibnamefont{and}
  \bibinfo{author}{\bibfnamefont{W.~H.} \bibnamefont{Press}},
  \bibinfo{journal}{J. Math. Phys.} \textbf{\bibinfo{volume}{14}},
  \bibinfo{pages}{7} (\bibinfo{year}{1973}).

\bibitem[{\citenamefont{Teukolsky}(1973)}]{Teukolsky:1973ha}
\bibinfo{author}{\bibfnamefont{S.~A.} \bibnamefont{Teukolsky}},
  \bibinfo{journal}{Astrophys. J.} \textbf{\bibinfo{volume}{185}},
  \bibinfo{pages}{635} (\bibinfo{year}{1973}).

\bibitem[{\citenamefont{Amaro-Seoane et~al.}(2007)}]{AmaroSeoane:2007aw}
\bibinfo{author}{\bibfnamefont{P.}~\bibnamefont{Amaro-Seoane}}
  \bibnamefont{et~al.}, \bibinfo{journal}{Class. Quant. Grav.}
  \textbf{\bibinfo{volume}{24}}, \bibinfo{pages}{R113} (\bibinfo{year}{2007}),
  \eprint{astro-ph/0703495}.

\bibitem[{\citenamefont{Yunes et~al.}(2009)\citenamefont{Yunes, Arun, Berti,
  and Will}}]{Yunes:2009yz}
\bibinfo{author}{\bibfnamefont{N.}~\bibnamefont{Yunes}},
  \bibinfo{author}{\bibfnamefont{K.~G.} \bibnamefont{Arun}},
  \bibinfo{author}{\bibfnamefont{E.}~\bibnamefont{Berti}}, \bibnamefont{and}
  \bibinfo{author}{\bibfnamefont{C.~M.} \bibnamefont{Will}}
  (\bibinfo{year}{2009}), \eprint{0906.0313}.

\bibitem[{\citenamefont{Mino et~al.}(1997)\citenamefont{Mino, Sasaki, Shibata,
  Tagoshi, and Tanaka}}]{Mino:1997bx}
\bibinfo{author}{\bibfnamefont{Y.}~\bibnamefont{Mino}},
  \bibinfo{author}{\bibfnamefont{M.}~\bibnamefont{Sasaki}},
  \bibinfo{author}{\bibfnamefont{M.}~\bibnamefont{Shibata}},
  \bibinfo{author}{\bibfnamefont{H.}~\bibnamefont{Tagoshi}}, \bibnamefont{and}
  \bibinfo{author}{\bibfnamefont{T.}~\bibnamefont{Tanaka}},
  \bibinfo{journal}{Prog. Theor. Phys. Suppl.} \textbf{\bibinfo{volume}{128}},
  \bibinfo{pages}{1} (\bibinfo{year}{1997}), \eprint{gr-qc/9712057}.

\bibitem[{\citenamefont{Yunes and Berti}(2008)}]{Yunes:2008tw}
\bibinfo{author}{\bibfnamefont{N.}~\bibnamefont{Yunes}} \bibnamefont{and}
  \bibinfo{author}{\bibfnamefont{E.}~\bibnamefont{Berti}},
  \bibinfo{journal}{Phys. Rev.} \textbf{\bibinfo{volume}{D77}},
  \bibinfo{pages}{124006} (\bibinfo{year}{2008}), \eprint{0803.1853}.

\bibitem[{\citenamefont{Poisson and Visser}(1995)}]{Poisson:1995sv}
\bibinfo{author}{\bibfnamefont{E.}~\bibnamefont{Poisson}} \bibnamefont{and}
  \bibinfo{author}{\bibfnamefont{M.}~\bibnamefont{Visser}},
  \bibinfo{journal}{Phys. Rev.} \textbf{\bibinfo{volume}{D52}},
  \bibinfo{pages}{7318} (\bibinfo{year}{1995}), \eprint{gr-qc/9506083}.

\bibitem[{\citenamefont{Poisson}(1993)}]{Poisson:1993vp}
\bibinfo{author}{\bibfnamefont{E.}~\bibnamefont{Poisson}},
  \bibinfo{journal}{Phys. Rev.} \textbf{\bibinfo{volume}{D47}},
  \bibinfo{pages}{1497} (\bibinfo{year}{1993}).

\bibitem[{\citenamefont{Flanagan and Hinderer}(2008)}]{Flanagan:2007ix}
\bibinfo{author}{\bibfnamefont{E.~E.} \bibnamefont{Flanagan}} \bibnamefont{and}
  \bibinfo{author}{\bibfnamefont{T.}~\bibnamefont{Hinderer}},
  \bibinfo{journal}{Phys. Rev.} \textbf{\bibinfo{volume}{D77}},
  \bibinfo{pages}{021502} (\bibinfo{year}{2008}), \eprint{0709.1915}.

\bibitem[{\citenamefont{Mazur and Mottola}(2004)}]{Mazur:2004fk}
\bibinfo{author}{\bibfnamefont{P.~O.} \bibnamefont{Mazur}} \bibnamefont{and}
  \bibinfo{author}{\bibfnamefont{E.}~\bibnamefont{Mottola}},
  \bibinfo{journal}{Proc. Nat. Acad. Sci.} \textbf{\bibinfo{volume}{101}},
  \bibinfo{pages}{9545} (\bibinfo{year}{2004}), \eprint{gr-qc/0407075}.

\bibitem[{\citenamefont{Cardoso
  et~al.}(2008{\natexlab{a}})\citenamefont{Cardoso, Pani, Cadoni, and
  Cavaglia}}]{Cardoso:2007az}
\bibinfo{author}{\bibfnamefont{V.}~\bibnamefont{Cardoso}},
  \bibinfo{author}{\bibfnamefont{P.}~\bibnamefont{Pani}},
  \bibinfo{author}{\bibfnamefont{M.}~\bibnamefont{Cadoni}}, \bibnamefont{and}
  \bibinfo{author}{\bibfnamefont{M.}~\bibnamefont{Cavaglia}},
  \bibinfo{journal}{Phys. Rev.} \textbf{\bibinfo{volume}{D77}},
  \bibinfo{pages}{124044} (\bibinfo{year}{2008}{\natexlab{a}}),
  \eprint{0709.0532}.

\bibitem[{\citenamefont{Cardoso
  et~al.}(2008{\natexlab{b}})\citenamefont{Cardoso, Pani, Cadoni, and
  Cavaglia}}]{Cardoso:2008kj}
\bibinfo{author}{\bibfnamefont{V.}~\bibnamefont{Cardoso}},
  \bibinfo{author}{\bibfnamefont{P.}~\bibnamefont{Pani}},
  \bibinfo{author}{\bibfnamefont{M.}~\bibnamefont{Cadoni}}, \bibnamefont{and}
  \bibinfo{author}{\bibfnamefont{M.}~\bibnamefont{Cavaglia}},
  \bibinfo{journal}{Class. Quant. Grav.} \textbf{\bibinfo{volume}{25}},
  \bibinfo{pages}{195010} (\bibinfo{year}{2008}{\natexlab{b}}),
  \eprint{0808.1615}.

\bibitem[{\citenamefont{Chirenti and Rezzolla}(2008)}]{Chirenti:2008pf}
\bibinfo{author}{\bibfnamefont{C.~B. M.~H.} \bibnamefont{Chirenti}}
  \bibnamefont{and} \bibinfo{author}{\bibfnamefont{L.}~\bibnamefont{Rezzolla}},
  \bibinfo{journal}{Phys. Rev.} \textbf{\bibinfo{volume}{D78}},
  \bibinfo{pages}{084011} (\bibinfo{year}{2008}), \eprint{0808.4080}.

\end{thebibliography}

\end{document}